# One Possible Reason for Double-Peaked Maxima in Solar Cycles: Is a Second Maximum of Solar Cycle 24 Coming?

A. Kilcik and A. Ozguc

*Akdeniz University, Faculty of Science, Department of Space Science and Technologies, 07058 Antalya, Turkey*

*Kandilli Observatory and Earthquake Research Institute, Bogazici University, 34684 Istanbul, Turkey*

Abstract: We investigate solar activity by focusing on double maxima in solar cycles and try to estimate the shape of the current solar cycle (Cycle 24) during its maximum. We analyzed data for Solar Cycle 24 by using Learmonth Solar Observatory sunspot group data since 2008. All sunspot groups (SGs) recorded during this time interval were separated into two groups: The first group includes small SGs [A, B, C, H, classes by the Zurich classification], and the second group consists of large SGs [D, E, and F]. We then calculated small and large sunspot group numbers, their sunspot numbers [SSN] and Zurich numbers [$Rz$] from their daily mean numbers as observed on the solar disk during a given month. We found that the temporal variations for these three different separations behave similarly. We also analyzed the general shape of solar cycles from Cycle 1 to 23 by using monthly International Sunspot Number [ISSN] data and found that the durations of maxima were about 2.9 years. Finally, we used ascending time and SSN relationship and found that the maximum of the Cycle 24 should be later than 2011. Thus, we conclude that i) one possible reason for a double maximum in solar cycles is the different behavior of large and small sunspot groups, and ii) a double maximum is coming for Solar Cycle 24.

1. Introduction

It is well known that solar activity is far from being constant. Variations in solar activity may be seen in many solar activity indicators, such as SSN, sunspot area, total solar irradiance, solar flares, *etc*. Due to their long coverage ($\cong$400 years), the SSN and sunspot area [SSA] are the most commonly used solar activity indicators, many of which are strongly correlated with each other (Hathaway, Wilson, and Reichmann, 2002; Deng *et al*., 2013, and references therein). SSN and SSA have been observed for centuries and it was determined that the solar cycle is not perfectly regular: cycle length and amplitude, ascending and descending times, the shape of maximums, vary from cycle to cycle (Usoskin and Mursula, 2003). The sunspot numbers are usually calculated as daily Wolf numbers:

$$Rz = k\,(10\,g + f) \qquad (1)$$

where *f* is the number of individual spots, *g* is the number of SGs, and *k* is a correction factor, that varies with location, observer, and instrumentation (Wolf, 1850; McIntosh, 1981; 1990). As shown in this equation, $R_z$ is strongly dependent on the number of observed groups and it does not have any weighting parameters for different size sunspots; A (a pore) and F (big evaluated spot) Zurich group sunspots have the same weight. Recent articles show that different groups of sunspots behave differently over a cycle (Kilcik *et al*., 2011; Lefevre and Clette, 2011; Javaraiah, 2012); Kilcik *et al*. (2011) separated sunspots into large and small SGs for the last four solar cycles and found that in general large SGs peaked about two years later than the small ones. It is well known that some sunspot cycles have twin peaks during their maxima. The existence of double peaks in a solar activity maximum was first noted by Gnevyshev (1963) for the coronal activity, and named as the Gnevyshev gap. Gnevyshev analyzed coronal intensity, prominences, and sunspots during Solar Cycle 18 (from 1953 to 1962) and concluded that the double maximum is valid for all cycles. Later it was found that the 11-year cycle consists of two distinct processes that are superimposed in time (Gnevyshev, 1967). For example, Cycle 23 had twin peaks and according to ISSN data; the cycle maximum occurred in the middle of 2000 (first maximum), but the other indicators peaked about two years later than the ISSN (second maximum). Thus, we may ask which is the real maximum and what is the reason for a double maximum during a cycle? Answering these questions will help to understand the background physics of the phenomena, and consequently it may increase the knowledge about Sun–Earth interactions. To answer some of these questions for the current solar cycle (Cycle 24); we focus on a comparison of two types of SGs, description of the duration of maxima of historical solar cycles, and the relations between ascending time and the largest smoothed monthly mean ISSN during a cycle maximum since 1755 (Cycle 1). Thus, we try to explain one possible reason for a double-peak maximum in a cycle and estimate the possible time of maximum of the current solar cycle.

2. Data, Methods and Analysis

All of the data used in this study are taken from National Geophysical Data Center (NGDC: ftp://ftp.ngdc.noaa.gov/STP/SOLAR_DATA/SUNSPOT_REGIONS/). Here, we analyzed three data sets from three points of view.

2.1. Behavior of SG Numbers and *Rz*

The modified Zurich classification SG data were used for comparing the temporal variation of large and small SGs and *Rz* since 2008. The SG classification data are collected by United States Air Force/Mount Wilson (USAF/MWL) Observatory. For the data used the USAF/MWL database includes measurements from the Learmonth Solar Observatory (22S, 114E), Holloman Solar Observatory (33N, 106W), and San Vito Solar Observatory (41N, 18E). We used the Learmonth station data as the principal data source, which has the best coverage for the investigated time interval. All Learmonth data gaps were filled with data from one of the other stations listed above, so that a nearly continuous time series was produced. The produced time series still had a gap around the cycle minimum, but this gap does not affect our results. For every SG in the data set, a classification according to the modified Zurich classification was assigned (McIntosh, 1990; Lefevre and Clette, 2011). It is possible to separate them into two distinct types. Here we used the same separation and data-preparation method described by Kilcik *et al*. (2011). In addition, we calculated the $R_z$ according to Equation (1) for each day and compared them. The temporal variation of SG numbers and calculated $R_z$ are given in Figure 1. As shown in this figure, large and small SG numbers and $R_z$ show almost the same behavior during a given temporal interval. From both data sets (12 months smoothed monthly data), the Solar Cycle 24 maximum occurred in February 2012. We would like to note that the maximum of this cycle (Cycle 24) starts in June 2011.

2.2 Duration of Solar Maxima

To find the average duration of the maximum of historical cycles (from Cycle 1 to Cycle 23) the ISSN data, which are taken from the NGDC web page as a monthly mean, were used. The monthly ISSN data is available back to 1700; according to Kane (2008), the quality of the data is reliable since 1848. Therefore, we used this part of the ISSN data as a reference and described the determination method of the first and second peak times as follows.

The largest smoothed ISSN data were used as a first parameter in our description from Cycle 1 to Cycle 23. As a second parameter, we took 15 % of the maximum values of the largest monthly ISSN for each corresponding cycle. Then, we subtracted them from the largest smoothed ISSN. Thus, we obtained the best SSN values that describe the smallest SSN of a maximum [hereafter $SSN_{limit}$], as a function of the maximum SSN of each cycle. We can express the description method as

$$SSN_{limit} = A - 0.15\ B \qquad (2)$$

where $SSN_{limit}$ is the smallest SSN that describes the beginning and end of maximum of a cycle, *A* is the largest smoothed monthly maximum SSN, and *B* is the monthly maximum SSN of the corresponding cycle. However, the monthly data may have strong small peaks in ascending and descending phases. Therefore, to smooth the fluctuations of monthly ISSN, the $SSN_{limit}$ values used in three steps running average data. Thus, the time of first and second peaks were described. Two examples are given in Figure 2.

As shown in this figure the duration of the maximum of Cycle 10 is about 2.42 years (from 1859.08 to 1861.50) and for Cycle 20, the value is about 3.75 years (1967 to 1970.75).

Finally, using Equation (2), we described the beginning and end of each cycle. In Table 1, some parameters of solar cycles from Cycle 1 to Cycle 23 are presented. There are several interesting points that we would like to address, regarding this table: i) Duration of solar maxima varies between 1.25 (Cycle 11) and 4.67 (Cycle 14), and the average value is 2.91 years. Thus, we may speculate that the solar activity has the strongest intensity for about 2.91 years for a cycle. ii) It is interesting to note that in general, the duration of maxima of weak cycles is longer than that of the strong/intense cycles; there are nine cycles that have 100 or less smoothed SSN and their average maximum duration is 3.35 years. iii) We would like to mention that most cycles have negative slope during their maxima. There are only four cycles that have positive – or zero – slopes, and they also have longer maximum duration. The average maximum duration of positive or zero slope cycles are 3.69 years. iv) The other interesting point is that a weak negative correlation ($r = 0.42$, $df = 21$, and $p < 0.05$) between duration of maxima and the largest smoothed sunspot numbers was determined (Figure 3).

2.3 Rise Time of a Cycle

It is known that there is a negative relation between the rise time and the largest smoothed monthly mean ISSN (Kilcik *et al*., 2009). They obtained a 0.82 negative correlation between two parameters by fitting them to the best linear function. The negative relationship between rise time to maximum and logarithmic maximum SSN was noted by Waldmeier (1935). Here, we also plotted the same two parameters in Figure 4 and fitted the best logarithmic function. Thus we obtained higher negative correlation ($r = 0.86$, $df = 21$, and $p < 0.001$) than the linear approximation. We derived Equation (3) between these parameters from the fit.

$$y = -2.648 \ln(x) + 16.663 \qquad (3)$$

where *x* and *y* indicate smoothed monthly mean SSN and the rise time to maximum for a cycle respectively. By using Equation (3) we tried to estimate the time of maximum for the current cycle (Cycle 24) by assuming the maximum monthly smoothed SSN will be 70 – 80 and found that the maximum will happen around the end of 2013 or the beginning of 2014.

3. Conclusions and Discussion

In this study, we tried to explain one possible reason for the double maximum in solar cycles and the shape of current solar cycle by using three approximations. The main findings of this study are as follows:

i) The temporal distributions of large and small SG numbers and $R_z$ show almost the same trend for the current cycle.

ii) The average duration of the maximum of historical solar cycles is about 2.91 years; in general, the duration of maximum of weak cycles are longer than the strong/intense cycles.

iii) In general, SSNs show a decreasing trend from the beginning of a maximum to the end during the maximum of a cycle. There are only four cycles that have positive – or zero – slope, and they also have longer maximum duration (3.69 years).

iv) There is a remarkable negative correlation (r = 0.86) between rise time to maximum and the smoothed monthly mean ISSN such that when the rise time to maximum getting shorter, the ISSN getting larger. By using Equation (3), we predict that the maximum of current solar cycle will occur around the end of 2013 or in the beginning of 2014 with 70 – 80 smoothed largest monthly averaged ISSN.

v) The four points mentioned above support that Cycle 24 will continue about one year more and we claim that at least two peaks will appear during the maximum. Our suggestion is that one possible reason for a double-peaked maximum in a solar cycle is the different behavior of large (complex) and small (simple) SGs; resulting from the existence of two different dynamo mechanisms.

We begin the discussion by pointing out that Kilcik *et al.* (2011) analyzed the temporal variation of small and large sunspot group (SG) numbers over the last four cycles and concluded that the large and small SG numbers behave differently during a cycle. It was determined that small SG numbers peaked at the same time with ISSN, which corresponds to the first maximum of a cycle, whereas large SG numbers peaked about two years later, corresponding to the second maximum. Kilcik *et al.* (2011) compared large and small SG numbers with some other solar activity indicators ($F_{10.7}$, facular area, SSA, and the maximum CME speed index) and found that the large groups describe all of these parameter better than small SG numbers do. Here, we calculated $R_z$ for large and small SGs by using of Equation (1) without a correction factor. As shown in Figure 1 both monthly total SG numbers and $R_z$ are following each other and there is an upturn at the end of the plots. Thus we may speculate that the maximum of this cycle is still continuing.

None of the solar cycles investigated have a sharp (less than 1.25 years) maximum during their maxima. The average duration in the maximum of historical cycles is about 2.91 years. We would like to underline that there is a weak negative relation ($r = 0.42$) between the duration in the maximum and the largest smoothed average SSN in the maximum of historical cycles. This indicates that if the amplitude of a cycle is smaller, it tends to the occurrence of longer duration in the maximum. The largest smoothed SSN measured is around 70 for the current cycle. If the above weak relation is right, the duration of the maximum of the current cycle should be longer than the average value (2.91 years). If the maximum of the current cycle occurred on November 2011, it should continue at least 2.91 years, so the Sun should have about 1.3 years more high activity.

Earlier we found that there is a remarkable negative correlation (r = 0.82) between rise time to maximum and the smoothed monthly mean ISSN, such that when the rise time to maximum becomes shorter, the SSN becomes larger (Pishkalo, 2008; Kilcik *et al.,* 2009). Here, we fitted the best logarithmic function to the same data and obtained a higher negative correlation coefficient ($r = 0.86$). This fitting function (Equation (3)) gave us the possible time of maximum depending on the largest smoothed monthly ISSN. The relationship between rise time to maximum and the largest SSN of a cycle has been used as a prediction tool for solar activity for a long time (Vitinskii, 1965; Du, 2011). Most of the recent articles predicted that the amplitude of the current solar cycle will be lower than 100 (Uzal, Piacentini, and Verdes, 2012; Passos, 2012; Kakad, 2011; Du, 2011, and references therein). From the current amplitude of Solar Cycle 24 (about 70) and results of the above mentioned predictions we assumed that the ISSN will be around 70 – 80 in the cycle maximum. Thus, we obtained the rising time to maximum as 5.1 – 5.4 years. It is known that Solar Cycle 24 started on December 2008, and at the time of writing [August 2013] only 4.67 years have passed from the minimum. From the combination of point 2 and 3 we may expect a longer maximum duration for current cycle and also positive – or zero – slope. Thus, we may speculate that solar activity will stay at around the same level relative to first maximum until the end of 2013.

We would like to emphasize that Gnevyshev (1963, 1967) reported that the first maximum of a cycle occurred because of active regions in all heliographic latitudes, while the second maximum arose due to the equatorial active regions. He concluded that the 11-year cycle consist of a partial superposition of these two distinct groups in time. Later Ferminella and Storini (1997) concluded that "structured maxima (two or three peaks) occur in each activity cycles as a result of dynamical effects superimposed on the quasi-periodic 11 years trend". They also concluded that the appearance of double peak is strongly related to complex events. All of these findings support our results. Kilcik *et al.* (2011) found that large (complex) and small (simple) SG numbers reach to their maxima in two different times of a maximum. These two different behaviors of large and small sunspot groups may be the result of two different dynamo mechanisms as found by Schatten (2009); i) a deep global dynamo at the base of the convection zone that it is responsible from the large SGs, ii) a local dynamo mechanism at work in the superficial layers on small scales. Combination of all the above points and discussions support the idea that one possible reason for a double peaked maximum in a solar cycle is the different behavior of large and small SGs. As can be seen in Figure 1, the two data sets follow each other quite well. As a result, we may expect that large SGs should reach their maximum later than the small ones do (Kilcik *et al*., 2011). Relationships between the largest smoothed monthly mean ISSN and cycle's duration and rise time to the maximum also strengthens this possibility.

Acknowledgement

The authors are grateful to an anonymous referee for suggestions and helpful comments. This study was supported by the Akdeniz University Research Fund.

References

Deng, L.H., Li, B., Zheng, Y.F., Cheng, X.M., 2013, *New Astron.* **23**, 1. DOI: 10.1016/j.newast.2013.01.004

Du, Z.: 2011, *Solar Phys.* **273**, 231. DOI: 10.1007/s11207-011-9849-8

Ferminella, F., Storini M.: 1997, *Astron. Astrophys.* **322**, 311.

Gnevyshev, M.N.: 1963, *Sov. Astron.* **7**, 311.

Gnevyshev, M.N.: 1967, *Solar Phys.* **1**, 107. DOI: 10.1007/BF00150306

Hathaway, D.H.,Wilson, R.M., Reichmann, E.J.: 2002, *Solar Phys*. **211**, 357. DOI: 10.1023/A:1022425402664

Javaraiah, J., *Solar Phys*. **281**, 827. DOI: 10.1007/s11207-012-0106-6

Kakad, B.: 2011, *Solar Phys*. **270**, 393. DOI: 10.1007/s11207-011-9726-5

Kane, R. P.: 2008, *Solar Phys*. **248,** 203. DOI: 10.1007/s11207-008-9125-8

Kilcik, A., Anderson, C.N.K., Rozelot, J.P., Ye, H., Sugihara, G., Ozguc, A.: 2009, *Astrophys. J*, **693**, 1173. DOI: 10.1088/0004-637X/693/2/1173

Kilcik, A., Yurchyshyn, V.B., Abramenko, V., Goode, P. , Ozguc, A., Rozelot, J.P., Cao, W.: 2011, *Astrophys. J*. **731**, 30. DOI: 10.1088/0004-637X/731/1/30

Lefevre, L., Clette, F.A.: 2011, *Astron. Astrophys*. **536**, L11. DOI: 10.1051/0004-6361/201118034

McIntosh, P.S.: 1981, In: Cram. L.E., Thomas, J.H. (eds.) *The Physics of Sunspots*, Sacramento Peak National Solar Observatory, **7**.

McIntosh, P.S.: 1990, *Solar Phys*. **125**, 251. DOI: 10.1007/BF00158405

Passos, D.: 2012, *Astrophys. J*. **744**, 172. DOI: 10.1088/0004-637X/744/2/172

Pishkalo, M.I.: 2008, *Kinemat. Phys. Celes. Bod.* **24**, 242. DOI: 10.3103/S0884591308050036

Schatten, K.: 2009, *Solar Phys*. **255**, 3. DOI: 10.1007/s11207-008-9308-3

Usoskin, I. G., Mursula, K.: 2003, *Solar Phys*. **218**, 319. DOI: 10.1023/B:SOLA.0000013049.27106.07

Uzal, L.C., Piacentini, R.D., Verdes, P.F.: 2012, *Solar Phys*. **279**, 551. DOI: 10.1007/s11207-012-0030-9

Vitinskii, Yu.I.: 1965, *Solar-Activity Forecasting*, [Israel Program for Scientific Translations Jerusalem], NASA-TT-F-289.

Waldmeier, M.: 1935, *Astron. Mitt. Zürich* **14**, 133.

Wolf, R.: 1850, *Astron. Mitt. Eidenössischen Sternwarte Zurich*, **1**, 27.

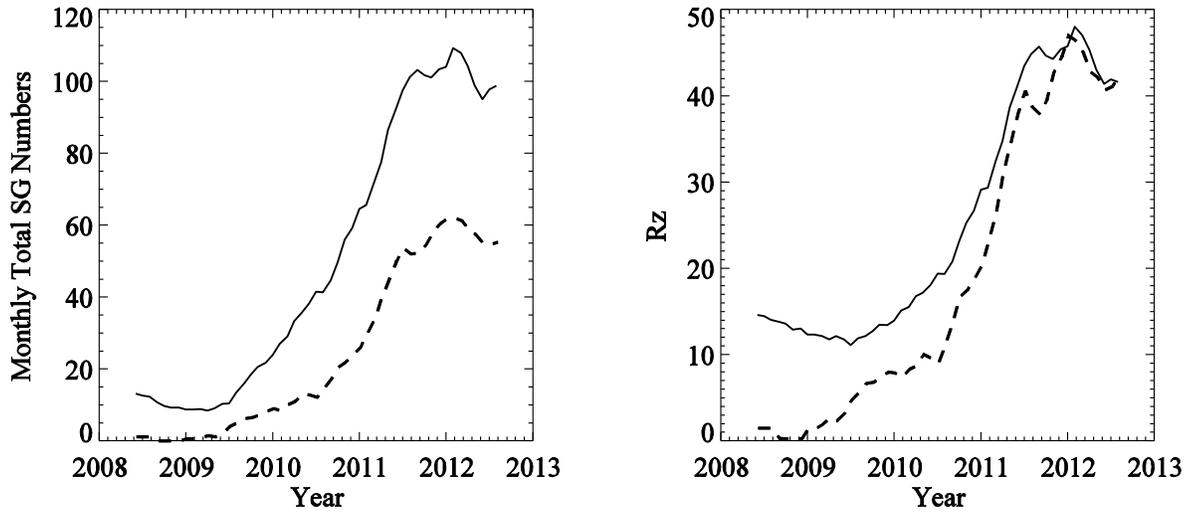

Figure 1. Temporal variation of monthly total large (dashed) and small (solid) SG numbers (left panel) and calculated daily average $Rz$ for each class (right panel). The data smoothed by a 12-step running-average smoothing method. Here we did not take into account the correction factor given in Equation (1).

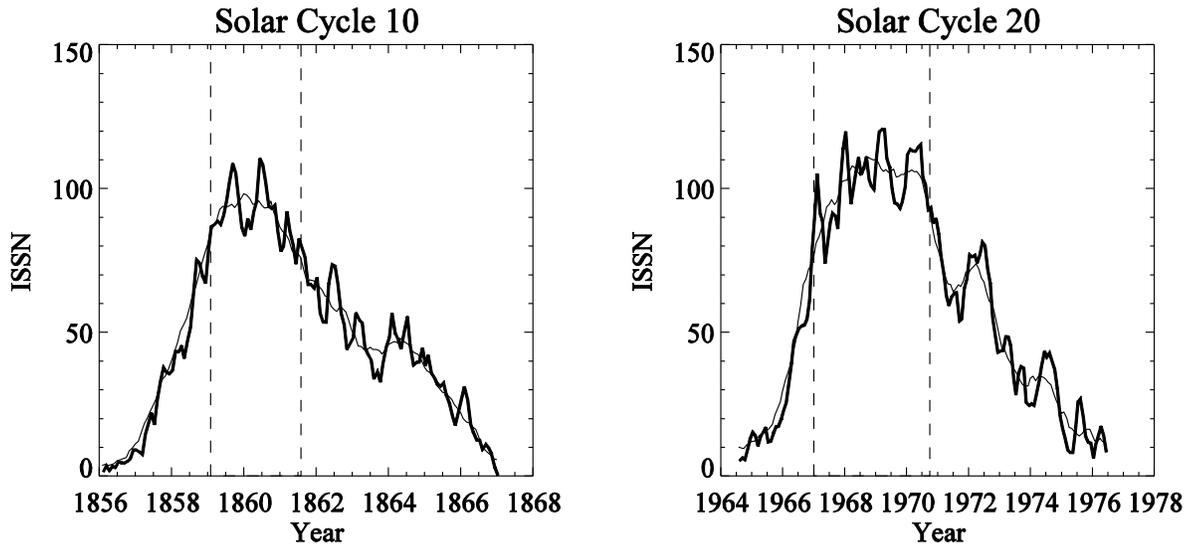

Figure 2. Two examples to illustrate description of beginning and end of maxima for Cycle 10 and Cycle 20. Vertical dashed lines show assumed reference maxima.

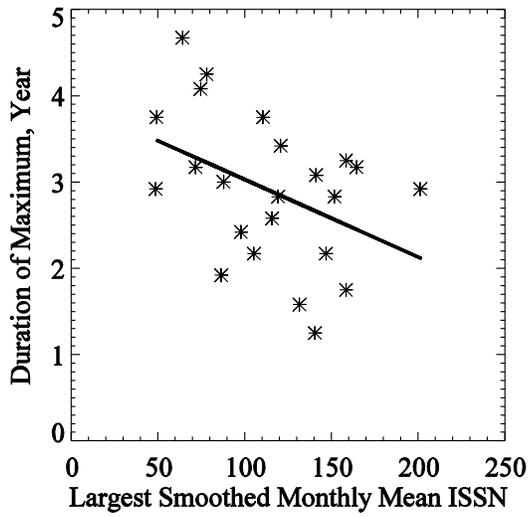

Figure 3. Relation between the largest smoothed monthly mean ISSN and the duration of solar maximum.

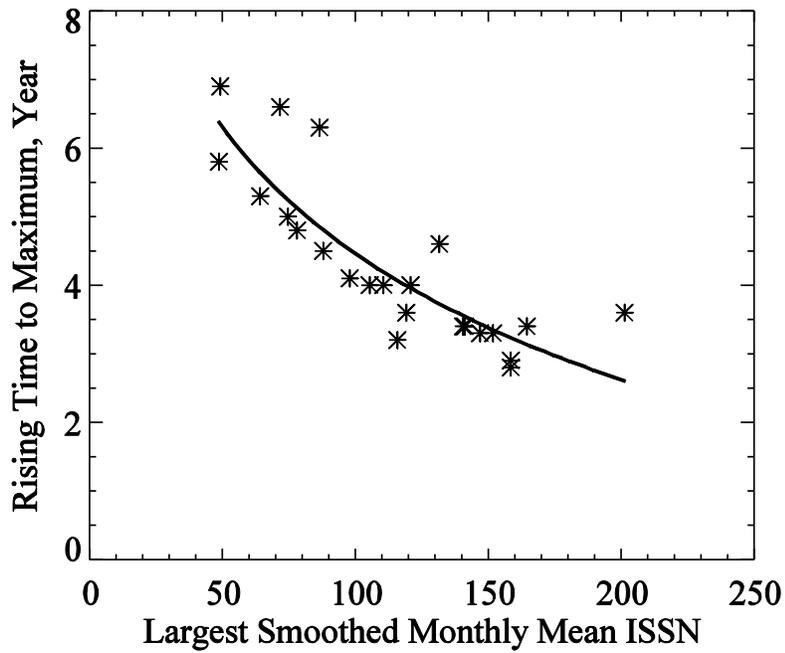

Figure 4. Relation between the largest smoothed monthly mean ISSN and the rise time to maximum.

Table 1. Some statistical parameters from Cycle 1 to Cycle 23.

| Cycle No. | Duration of Maximum [Years] | Largest Smoothed Monthly Mean | Largest Monthly Maximum SSN | $SSN_{limit}$ | Slope | Rise to Max [Years] | Cycle Length [Years] |
|---|---|---|---|---|---|---|---|
| 1 | 1.92 | 86.50 | 107.20 | 70.42 | -1.5 | 6.3 | 11.3 |
| 2 | 2.58 | 115.80 | 158.20 | 92.07 | -1.2 | 3.2 | 9.0 |
| 3 | 1.75 | 158.50 | 238.90 | 122.67 | -0.9 | 2.9 | 9.2 |
| 4 | 3.08 | 141.20 | 174.00 | 115.10 | -0.4 | 3.4 | 13.6 |
| 5 | 3.75 | 49.20 | 62.30 | 39.86 | 0.1 | 6.9 | 12.3 |
| 6 | 2.92 | 48.70 | 96.20 | 34.27 | -0.4 | 5.8 | 12.7 |
| 7 | 3.17 | 71.70 | 106.30 | 55.76 | 0.0 | 6.6 | 10.6 |
| 8 | 2.17 | 146.90 | 206.20 | 115.97 | -0.3 | 3.3 | 9.6 |
| 9 | 1.58 | 131.60 | 180.40 | 104.54 | -0.3 | 4.6 | 12.5 |
| 10 | 2.42 | 97.90 | 116.70 | 80.40 | -0.4 | 4.1 | 11.2 |
| 11 | 1.25 | 140.50 | 176.00 | 114.10 | -0.6 | 3.4 | 11.7 |
| 12 | 4.08 | 74.60 | 95.80 | 60.23 | 0.0 | 5.0 | 10.7 |
| 13 | 3.00 | 87.90 | 129.20 | 68.52 | -0.2 | 4.5 | 12.1 |
| 14 | 4.67 | 64.20 | 108.20 | 47.97 | -0.1 | 5.3 | 11.9 |
| 15 | 2.17 | 105.40 | 154.50 | 82.23 | -1.6 | 4.0 | 10.0 |
| 16 | 4.25 | 78.10 | 108.00 | 61.90 | -0.1 | 4.8 | 10.2 |
| 17 | 2.83 | 119.20 | 165.30 | 94.41 | -0.7 | 3.6 | 10.4 |
| 18 | 2.83 | 151.80 | 201.30 | 121.61 | -0.6 | 3.3 | 10.1 |
| 19 | 2.92 | 201.30 | 253.80 | 163.23 | -0.2 | 3.6 | 10.6 |
| 20 | 3.75 | 110.60 | 135.80 | 90.23 | 0.3 | 4.0 | 11.6 |
| 21 | 3.17 | 164.50 | 188.40 | 136.24 | -0.4 | 3.4 | 10.3 |
| 22 | 3.25 | 158.50 | 200.30 | 128.46 | -0.5 | 2.8 | 9.7 |
| 23 | 3.42 | 120.80 | 170.10 | 95.29 | -0.1 | 4.0 | 12.5 |